\documentstyle[prl,aps,twocolumn,epsfig,amstext]{revtex}
\begin{document}
\def\d{$^{\circ}$}
\def\am{{$a_{\mu}$} }
\def\g2{{$(g-2)$} }
\def\ms{{$\mu$s}}
\def\be{\begin{equation}}
\def\ee{\end{equation}}
\def\bea{\begin{eqnarray}}
\def\eea{\end{eqnarray}}
\def\bc{\begin{center} }
\def\ec{\end{center} }
%\baselineskip=10pt

% A useful Journal macro
\def\Journal#1#2#3#4{{#1} {\bf #2}, #3 (#4)}

% Some useful journal names
\def\NCA{\em Nuovo Cimento}
\def\NIM{\em Nucl. Instrum. Methods}
\def\NIMA{{\em Nucl. Instrum. Methods} A}
\def\NPB{{\em Nucl. Phys.} B}
\def\NPS{Nucl. Phys. B (Proc. Suppl.)}
\def\PLB{{\em Phys. Lett.}  B}
\def\PRL{\em Phys. Rev. Lett.}
\def\PRD{{\em Phys. Rev.} D}
\def\PRA{{\em Phys. Rev.} A}
\def\ZPC{{\em Z. Phys.} C}
\def\Met{{\em Metrologia}}
\def\EPC{{\em Eur. Phys. J.} C}
\def\RMP{\em Rev. Mod. Phys.}
\def\RPP{\em Rep. Prog. Phys.}
% Some other macros used in the sample text
\def\st{\scriptstyle}
\def\sst{\scriptscriptstyle}
\def\mco{\multicolumn}
\def\epp{\epsilon^{\prime}}
\def\vep{\varepsilon}
\def\ra{\rightarrow}
\def\ppg{\pi^+\pi^-\gamma}
\def\vp{{\bf p}}
\def\ko{K^0}
\def\kb{\bar{K^0}}
\def\al{\alpha}
\def\ab{\bar{\alpha}}
\def\CPbar{\hbox{{\rm CP}\hskip-1.80em{/}}}%temp replacement due to no font
\bibliographystyle{unsrt}    % for BibTeX - sorted numerical labels

\title{Improved Measurement of the Positive Muon Anomalous Magnetic 
Moment$^{\dag}$}

\author{ Muon $(g-2)$ Collaboration\\ 
H.N. Brown$^2$, G. Bunce$^2$, R.M. Carey$^1$,  P. Cushman$^{10}$, 
 G.T. Danby$^2$,  P.T. Debevec$^7$,  H. Deng$^{12}$, W.Deninger$^7$,
 S.K. Dhawan$^{12}$, V.P. Druzhinin$^3$,  L. Duong$^{10}$,  
W. Earle$^1$, E. Efstathiadis$^1$,  G.V. Fedotovich$^3$,  F.J.M. Farley$^{12}$,
S. Giron$^{10}$,
F. Gray$^7$, M. Grosse-Perdekamp$^{12}$, A. Grossmann$^6$, U. Haeberlen$^8$, 
M.F. Hare$^1$, E.S. Hazen$^1$, D.W. Hertzog$^7$, V.W. Hughes$^{12}$, 
M. Iwasaki$^{11}$,   K. Jungmann$^6$,
D. Kawall$^{12}$  M. Kawamura$^{11}$, B.I. Khazin$^3$, 
J. Kindem$^{10}$, F. Krienen$^1$, I. Kronkvist$^{10}$, 
R. Larsen$^2$, Y.Y. Lee$^2$, I. Logashenko$^{1,3}$, R. McNabb$^{10}$, 
W. Meng$^2$, J. Mi$^2$, J.P. Miller$^1$,  W.M. Morse$^2$, C.J.G. 
Onderwater$^7$, 
Y. Orlov$^4$,  C. \"Ozben$^2$, C. Polly$^7$,
C. Pai$^2$, J.M. Paley$^1$, J. Pretz$^{12}$,  R. Prigl$^2$, 
G. zu Putlitz$^6$, S.I. Redin$^{12}$, 
O. Rind$^1$, B.L. Roberts$^1$, N. Ryskulov$^3$, S. Sedykh$^7$, 
Y.K. Semertzidis$^2$, Yu.M. Shatunov$^3$,  E. Solodov$^3$, M. Sossong$^7$, 
A. Steinmetz$^{12}$,
L.R. Sulak$^1$, C. Timmermans$^{10}$,  A. Trofimov$^1$, D. Urner$^7$, 
P. von Walter$^6$,  D. Warburton$^2$, D. Winn$^5$, 
A. Yamamoto$^9$, D. Zimmerman$^{10}$
\\
$^1$Department of Physics, Boston University, Boston, MA 02215, USA\\
$^2$Brookhaven National Laboratory, Upton, NY 11973, USA\\
$^3$Budker Institute of Nuclear Physics, Novosibirsk, Russia\\
$^4$Newman Laboratory, Cornell University, Ithaca NY 14853, USA\\
$^5$Fairfield University, Fairfield, CT 06430, USA\\
$^6$Physikalisches Institut der Universit\"at Heidelberg, 69120
Heidelberg, Germany\\
$^7$Department of Physics, 
University of Illinois at Urbana-Champaign, IL 61801,
USA\\
$^8$MPI f\"ur Med. Forschung, 69120 Heidelberg, Germany\\
$^9$KEK, High Energy Accelerator Research Organization,Tsukuba, Ibaraki
305-0801, Japan\\
$^{10}$Department of Physics, University of Minnesota, Minneapolis, MN 55455, 
USA\\
$^{11}$Tokyo Institute of Technology, Tokyo, Japan\\
$^{12}$Department of Physics, Yale University, New Haven, CT 06511, USA\\
$^{\dag}${\bf To be published in Phys. Rev. D62, RAPID COMMUNICATIONS}
}

\date{\today}
\maketitle
\newpage
%
% abstract text goes here
%
\begin{abstract} 
A new measurement of the positive
muon's anomalous magnetic moment has been made at the
Brookhaven Alternating Gradient Synchrotron using the direct injection of
polarized muons into the superferric storage ring.
The angular frequency difference $\omega_{a}$ between the angular spin
precession frequency $\omega_{s}$ and the angular orbital frequency
$\omega_{c}$ is measured as well as the free proton NMR frequency 
$\omega_{p}$.
These determine  
  $R$ = $\omega_{a}/\omega_{p}$ = 3.707~201(19) $\times 10^{-3}$.
With $\mu_{\mu}/\mu_{p}$ = 3.183~345~39(10) this gives $a_{\mu^{+}}$
= 11~659~191(59)$ \times 10^{-10}$ ($\pm 5$ ppm), in good agreement with the
previous CERN and BNL measurements for $\mu^{+}$ and $\mu^{-}$,
and with  the standard model prediction.

\end{abstract}

\noindent pacs 14.60.Ef

\begin{center}

\end{center}

In this brief note we present a new measurement of the anomalous $g$-value
$a_{\mu}=(g-2)/2$ of the positive muon from the Brookhaven Alternating Gradient
Synchrotron (AGS) experiment E821.  We previously reported a result based on
data collected during 1997.\cite{carey} In that work, as in the CERN
measurements,\cite{cern3,fp} pions were injected 
into the storage ring, and approximately 25 parts per million (ppm)  of
the daughter muons from pion decay were stored.
In August 1998  a fast muon kicker was commissioned,
which permitted the direct injection of muons
into the storage ring.

Except for the use of muon injection, many of the experimental aspects are
the same as described in reference\cite{carey}.
However, important improvements were implemented.  These included
better stability and homogeneity of the storage ring magnetic field,
improved
stability of the positron detection system, and extended capacity of the 
data acquisition system.
For the 1998 run, the AGS contained six proton bunches, each with 
a maximum intensity of about 
$7\times 10^{12}$, with one bunch extracted every 33 ms. 
The 3.1 GeV/c positive muon beam was formed from decays of a secondary
pion beam which was 1.7\% higher in momentum, thus providing a 
muon polarization of about 95\%.
The pion decay channel
consists of a 72 m long straight  section of the
secondary beamline.
The muons were selected
at a momentum slit, where the (higher energy) pions were  directed
into a beam dump. The beam composition entering the storage ring was measured
with a threshold Cerenkov counter filled with isobutane.  By stepping
the pressure from zero to 1.2 atm, the threshold for Cerenkov light
from $e^+$, then $\mu^+$ and finally $\pi^+$ was crossed.  The beam 
was found to consist of equal parts of positrons, muons
and pions, consistent with Monte Carlo predictions.  While this measurement 
was not sensitive to the proton content of the beam, calculations 
predict it to be approximately one third of the pion flux. 
The flux incident on the
storage ring was typically  $2 \times 10^{6}$ for each
proton bunch. 

The 10 mrad kick needed to put the muon beam 
onto a stable orbit was achieved
with a peak current of 4100 A and a half period of 400 ns.
   Three
pulse-forming networks powered three identical 1.7 m long one-loop
kicker sections consisting of 95 mm high
parallel plates on either side of the beam.
The current pulse was formed by an 
under-damped LCR circuit.
The kicker plate geometry and composition were
chosen to minimize eddy currents.
The residual eddy current effect on the total
field seen by the muons was less than 0.1 ppm 20 $\mu$s after injection. 
The time-varying magnetic field from the eddy currents was calculated with
the program OPERA\cite{opera} and was measured in a full-size straight
prototype vacuum chamber with the use of the Faraday effect.\cite{Faraday}
Since the muons circulate in
149 ns, they were kicked several times before the kicker pulse died out.  

About $10^{4}$ muons were stored in the ring per proton bunch.
With muon injection,
the number of detected positrons per hour was increased by an
order of magnitude over the pion-injection method employed previously.
Furthermore,  the injection related background (flash)
in the positron detectors was
reduced by a factor of about 50,
since most of the pions were removed from the beam before
entering the storage ring.

For polarized muons moving in a uniform 
magnetic field $\vec B$
perpendicular to the muon spin direction and to
the plane of the orbit
and with an electric quadrupole field $\vec E$,
which is used for
vertical focusing\cite{cern3,quads},  
the  angular frequency difference, $\omega_a$,
between the spin precession frequency $\omega_s$ and the cyclotron frequency
$\omega_c$, is given by
\be
\vec \omega_a = - {e \over m }\left[ a_{\mu} \vec B -
\left( a_{\mu}- {1 \over \gamma^2 - 1}\right)
\vec \beta  \times \vec E \right].
\label{eq:omega}
\ee
The dependence of $\omega_a$ on the electric field 
is eliminated by storing
muons with the ``magic'' $\gamma$=29.3, which 
corresponds to a muon momentum
$p$ = 3.09 GeV/$c$.  Hence measurements of $\omega_a$ and of $B$
determine $a_\mu$. 
At the magic gamma, the muon lifetime  is 
$\gamma \tau =64.4$ \ms\  and
the \g2 precession period is 4.37 \ms.
With a field of 1.45 T in
our storage ring,\cite{danby} 
the central orbit radius is 7.11 m.

\begin{figure}[h]
%epsfxsize = 3.4 in
\epsfysize = 3.2 in
\hfil
\epsffile {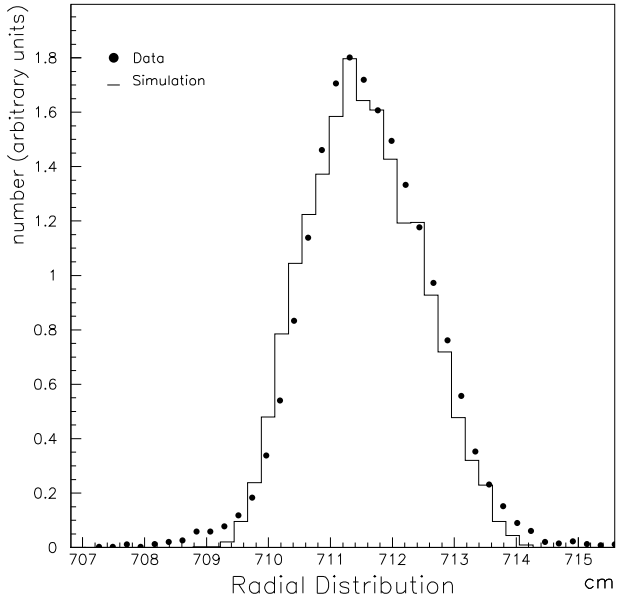}
\hfil
%\vskip -0.5 cm
\caption[6]{{The equilibrium radius
distribution calculated using the tracking code
(histogram) and obtained from an analysis of the beam debunching at early
times (points).
}
\label{fg:fastrot}}
\end{figure}

The magnetic field in  Eq. \ref{eq:omega}
is the average over
the muon distribution.  We
obtained the equilibrium radius distribution
by determining the distribution of
rotation frequencies  in  the  ring  from  the  time  spectra  of  decay
positrons.\cite{carey}  
The distribution, reproduced with a tracking code, was found
to be 3 mm toward the outside of the central storage region. This offset
was caused by the mode of operating the kicker.  The calculated and
measured radial distributions are shown in Fig. \ref{fg:fastrot}.

The  magnetic  field  seen  by  the  muon distribution was calculated 
by tracking a sample  of muons in
software through the  field  map measured by NMR, 
and  by averaging the field values.
The resulting
average corresponds within 0.02 ppm to the  field  value
taken at the beam center and averaged over azimuth.   We used the latter
to account for variations with time, and to obtain the present result.

\begin{figure}[h]
\epsfxsize = 3.4 in
\epsfysize = 3.4 in
\hfil
\epsffile {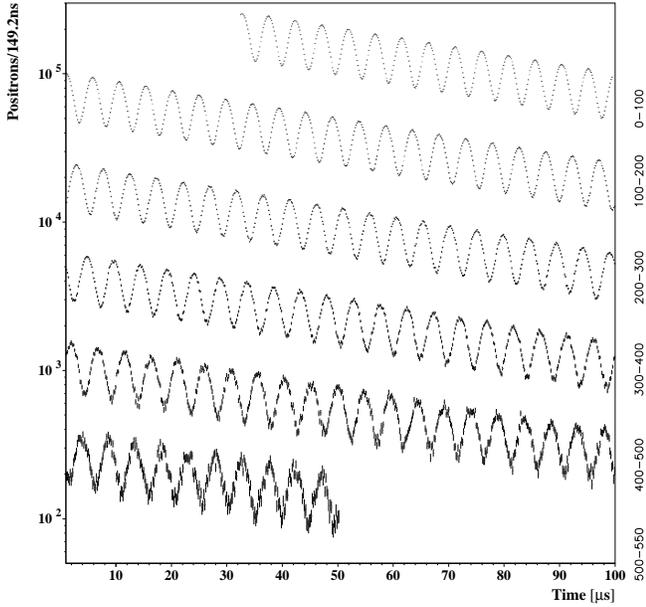}
\hfil
%\vskip -0.5 cm
\caption[6]{{The positron time spectrum obtained with muon injection
for $E>1.8$ GeV.  These data represent 84 million positrons.
}
\label{fg:eparft}}
\end{figure}

Positrons from the in-flight decay 
$\mu^+ \rightarrow e^{+}\nu_{e}\bar{\nu}_{\mu}$ were 
detected with 24 Pb-scintillating
fiber calorimeters\cite{calo} 
placed symmetrically around the inside of
the storage ring.  Twenty-one of these 
detectors were used in the present analysis.
The observed positron time spectrum shown in Fig. \ref{fg:eparft}
was adequately represented
by\cite{cern3,fp}
\be
N_{0}(E)e^{-t/\gamma\tau}[1+A(E)cos(\omega_{a}t+\phi(E))].
\label{eq:fivep} 
\ee
The normalization constant $N_0$ depends on the energy threshold, $E$,
placed upon the positrons. 
The (integral) asymmetry 
$A$ depends on $E$ and on
the beam polarization. The fractional statistical error on $\omega_{a}$ is
proportional to $A^{-1}N^{-1/2}_{e}$, where $N_{e}$ is the number of decay
positrons detected above threshold.  For an energy threshold 
of 1.8 GeV where $N_eA^2$ is maximum,
$A$ was found to be 0.34 on average.

As in Ref. \cite{carey}, 
the photomultiplier tubes were gated off before injection.
With the reduced flash associated with muon injection, it was possible to
begin counting as soon as 5 $\mu$s after injection in the region of the ring
270$^{\circ}$ around from the injection point, and 35 $\mu$s in the 
injection region.  Data from the detectors gated on at 5 $\mu$s were used
in the rotation frequency analysis mentioned above.  The fit to 
Eq. \ref{eq:fivep} was begun after scraping\cite{carey} was completed
and when the photomultiplier outputs from the 21 detectors 
used in this analysis were  stable, 25-40 $\mu$s after injection.
Time histograms were formed for each detector.  These data
were analyzed separately and
the resulting values for $\omega_a$
were in good agreement ($\chi^{2}/\nu$ = 17.2/20).

Values for $\omega_{a}$ and $\omega_{p}$, the free proton NMR angular 
frequency in the storage-ring
magnetic field,
were determined separately and independently. 
Thereafter the frequency
ratio  $R = \omega_a/\omega_p$ was determined. 
A correction of $+0.9$ ppm was added to $R$ to account for the effects\cite{fp}
of the electric field and the muon vertical betatron oscillations on
$\omega_a$.
We obtain  $R= 3.707\ 201 (19) \times 10^{-3}$,
where the 5 ppm error includes a 1 ppm systematic error discussed below.

\begin{figure}[p]
\epsfysize = 9. cm
\begin{center}
\epsffile{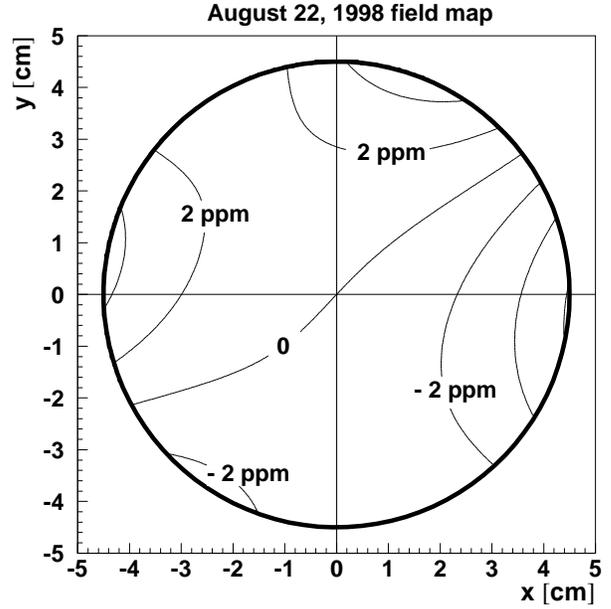}
\end{center}
%\vskip -0.5 cm
\caption[6]{{A magnetic field profile averaged over azimuth.
The circle encloses the muon storage region of 4.5 cm radius.
The contours represent 2 ppm changes in the
field.
}
\label{fg:field}}
\end{figure}

Since the 1997 run, a substantial reduction in the overall systematic
error has been achieved.
As regards $\omega_p$,
the stability of the magnetic field has been improved
by thermal insulation of the magnet and by NMR feedback control to the main
magnet power supply.  The field homogeneity has been improved by additional
shimming with iron shims near the intersections of pole pieces, and
the surface 
coils around the ring on the pole faces have been used to compensate on
average for the higher multipoles in the magnet.  Additional shimming
was also done using the iron wedges
placed in the air gap separating the high quality low carbon
pole piece steel  from the
yoke steel.
This shimming produced a field which, when averaged over the azimuth, was
uniform to within $\pm$ 4 ppm, as is shown in Fig. \ref{fg:field}.
The knowledge
of the muon distribution in the ring obtained as indicated above allows us to
determine the average field $<B>$ seen by the muons.  The uncertainty in
$<B>$ is $\pm 0.5$ ppm.

The other systematic errors are associated with the determination of 
$\omega_a$ from the positron data.  These arise principally from pile-up,
and from AGS ``flashlets''.
A pile-up error occurs when two pulses overlap within the time resolution
of about 5 ns and are incorrectly identified as one, which then gives
incorrect times and energies for the positrons.
Pile-up is estimated to produce an 
effect on $\omega_a$ of less than 0.6 ppm, which we conservatively take as
an error estimate.  Occasionally under unstable conditions, the AGS was 
observed to extract beam during our data collection period of 600 
$\mu$s which caused a background in our calorimeters (``flashlets'').
We conservatively estimate that this effect on $\omega_a$ in the 1998
data sample is less than 0.5 ppm.  Smaller errors arise from the 
details of the fitting procedure, rate dependent timing shifts, 
gain changes in the photomultipliers,
uncertainties about the radial electric field and the vertical betatron
motion, and from muon losses.

Altogether, the systematic errors on $\omega_a$ and $\omega_p$ added in
quadrature are less than 1 ppm.

The anomalous magnetic moment is obtained from the frequency ratio $R$ by
\be
a_{\mu^+} = { R \over { \lambda - R} } = 11\ 659\ 191 (59) \times 10^{-10}
\ee
in which $\lambda = \mu_{\mu}/\mu_p = 3.183\ 345\ 39(10)$\cite{pdg,lambda}.
This new result is in good agreement with the mean of the CERN
measurements
for $a_{\mu^+}$ and $a_{\mu^-}$\cite{fp,pdg}, and our previous measurement
of   $a_{\mu^+}$\cite{carey}.
 Assuming CPT symmetry,
the weighted mean of the four measurements
gives a new world average of
\be 
a_{\mu} = 11\ 659\ 205 (46) \times 10^{-10} \qquad (\pm 4\  {\rm ppm}),
\ee
($\chi^2/\nu = 2.7 / 3$).

The theoretical value of \am in the standard model has its dominant 
contribution from 
quantum electrodynamics but the strong and weak interactions contribute as
well.
The standard model value is 
$a_{\mu} ({\rm SM}) \  =\  11\ 659\ 163 (8) \times 10^{-10}$
($\pm 0.7$ ppm), where the error is dominated by the uncertainty in the
lowest order hadronic vacuum polarization.\cite{kinhughes}

\begin{figure}[htbp]
\epsfxsize = 2.8in
\epsfysize = 3.in
\hfil
\epsffile {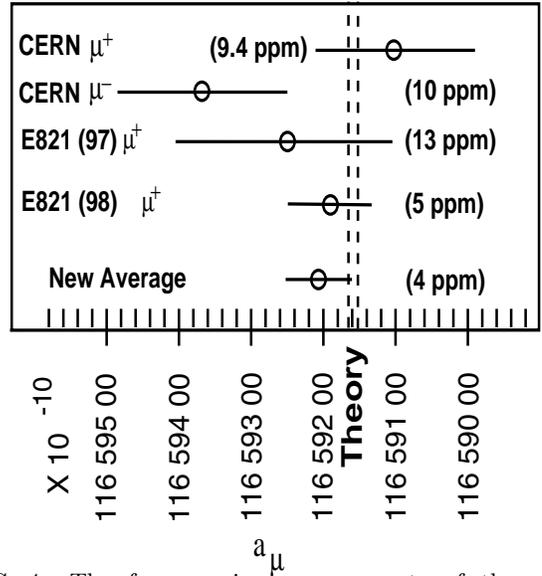}
\hfil
\caption[6]{{The four precise measurements of the muon anomalous magnetic
moment and their weighted average.  The $1\sigma$ region allowed by the 
standard model (see text) is indicated by the dashed lines. 
}
\label{fg:comp}}
\end{figure}

In Fig. \ref{fg:comp} the
four precise measurements of $a_{\mu}$ and their average
are shown, along with the standard model prediction.
 The weighted mean of the experimental results agrees
with the standard model, 
with 
\be
a_{\mu}{(\rm Expt.)} - a_{\mu}{(\rm Theory)} =
(42 \pm 47) \times 10^{-10} 
\ee
or equivalently $ (3.6 \pm 4.0)$  ppm.
This agreement of theory and experiment further constrains new
physics beyond the standard model.\cite{newphys1,newphys2}

Data collected in early 1999 should give a statistical error of about
1 ppm, with a systematic error below 1 ppm.
There is substantial activity at the $e^{+}e^{-}$ colliders of
Novosibirsk\cite{novosib} and
Beijing,\cite{bes} to measure $\sigma(e^+e^- \rightarrow {\rm hadrons})$, 
and these new data will further 
improve our knowledge
of the hadronic contribution and thereby the standard model value of 
$a_{\mu}$. 

We thank T.B.W. Kirk, D.I. Lowenstein, P. Pile and 
the staff of the BNL AGS for the strong support they have given this 
experiment.
We also thank C. Coulsey, G. De Santi and J. Sinacore, for their 
contributions to the preparation and running of the experiment.
This work was supported in part by the U.S. Department of Energy, the U.S.
National Science Foundation, the German Bundesminister f\"ur Bildung und
Forschung, The Russian Ministry of Science
and the US-Japan Agreement in High Energy Physics. A. Steinmetz
acknowledges support by the Alexander von Humboldt Foundation.

\vskip0.5in

\bigskip

\begin{thebibliography}{99}

%
%  list bibliography in order,  \bibitem {label}  ...... with period at end
%

\bibitem{carey}R.M. Carey et al., $(g-2)$ Collaboration,
\Journal{\PRL}{82}{1632}{1999}.


\bibitem {cern3} J. Bailey et al., \Journal{\NPB}{150}{1}{1979}.


\bibitem {fp} F.J.M. Farley and E. Picasso in {\em Quantum Electrodynamics},
ed. T. Kinoshita, (World Scientific, Singapore, 1990), p. 479.

\bibitem{opera} Vector Fields Limited, 24 Bankside, Kidlington,
Oxford OX5 1JE,  England.

\bibitem{Faraday} Muon $(g-2)$ note \#286  {\it The $(g-2)$ Muon Kicker: 
Design and Status}, E. Efstathiadis et al., (1997) and
E. Efstathiadis et al., `` The Muon $(g-2)$ Muon Kicker'', to be 
submitted  to {Nucl. Instrum. Methods}.

\bibitem{quads} Y.K. Semertzidis, et al. ``The Brookhaven Muon g-2 
Storage Ring High Voltage Quadruples'', to be submitted to Nucl. Inst. 
and Methods.

\bibitem{danby} G.D. Danby, et al., {\em Nucl. Instrum. Methods} 
in press (2000).

\bibitem {calo} S. Sedykh, et al., {\em Nucl. Instrum. Methods}, 
in press (2000).

\bibitem {pdg} Particle Data Group, \Journal{\EPC}{3}{1}{1998}.

\bibitem {lambda} W. Liu, et al., \Journal{\PRL}{82}{711}{1999}.

\bibitem{kinhughes} V.W. Hughes and T. Kinoshita, 
\Journal{\RMP}{71}{S133}{1999}.

\bibitem{newphys1} T. Kinoshita and W.J. Marciano in
{\em Quantum Electrodynamics} (Directions in High Energy Physics, Vol. 7),
ed. T. Kinoshita, (World Scientific, Singapore, 1990), p. 419.

\bibitem{newphys2}A. Czarnecki and W. Marciano, 
\Journal{\NPS}{76}{245}{1999}, and references therein.

\bibitem{novosib}R.R. Akhmetshin et al., (CMD2 Collaboration),
\Journal{\PLB}{475}{190}{2000} and references therein.

\bibitem{bes}Zhengguo Zhao, (arXiv:hep-ex/0002025 v2)
Proceedings for Lepton-Photon 99 to be
published  and
 J.Z. Bai, et al., (BES Collaboration), 
\Journal{\PRL}{84}{594}{2000}.

\end{thebibliography}
\end{document}